\documentclass[12pt, a4paper]{article}
\pdfoutput=1
\usepackage{a4wide}
\usepackage{amsfonts}
\usepackage{amscd}
\usepackage{amssymb}
\usepackage{amsmath}
\usepackage{epsfig}
\usepackage{latexsym}
\usepackage{mathtools}
\usepackage{stmaryrd}
\usepackage{hyperref}
\usepackage{setspace}
\hypersetup{
    colorlinks=true,
    linkcolor=black,
    citecolor=black,
    filecolor=black,
    urlcolor=black,
}

\def \be  {\begin{equation}}
\def \ee  {\end{equation}}
\def \ba  {\begin{eqnarray}}
\def \ea  {\end{eqnarray}}
\def \bb  {\begin{thebibliography}}
\def \eb  {\end{thebibliography}}
\def \lab #1 {\label{#1}}

\newcommand\cA{\mathcal{A}}
\newcommand\cB{\mathcal{B}}
\newcommand\cC{\mathcal{C}}
\newcommand\cD{\mathcal{D}}

\newcommand\cM{\mathcal{M}}
\newcommand\cN{\mathcal{N}}
\newcommand\cO{\mathcal{O}}
\newcommand\cP{\mathcal{P}}

\newcommand\cY{\mathcal{Y}}
\newcommand\cZ{\mathcal{Z}}
\newcommand\al{\alpha}
\newcommand\dal{\dot\alpha}

\newcommand\CP {\mathbb{CP}}

\newcommand\lb{\lambda}
\newcommand\tlb{\tilde{\lambda}}
\newcommand\teta{\tilde{\eta}}
\newcommand\la{\langle}
\newcommand\ra{\rangle}

\newcommand\tPsi{\widetilde{\Psi}}

\newcommand\tP{\widetilde{\mathcal{P}}}
\newcommand\tQ{\widetilde{\mathcal{Q}}}

\newcommand{\captionfonts}{\small}

\makeatletter 
\long\def\@makecaption#1#2{%
\vskip\abovecaptionskip
\sbox\@tempboxa{{\captionfonts #1: #2}}%
\ifdim \wd\@tempboxa >\hsize
  {\captionfonts #1: #2\par}
 \else
   \hbox to\hsize{\hfil\box\@tempboxa\hfil}%
  \fi
  \vskip\belowcaptionskip}
\makeatother

\begin{document}

\thispagestyle{empty}
\vskip1.5truecm
\begin{center}
\vskip 0.2truecm 

{\Large\bf New Formulae for Gravity Amplitudes:}
\vskip 3truemm
{\Large\bf  Parity Invariance and Soft Limits}
\vskip 1truecm
{\bf Mathew Bullimore}
\vskip 0.5truecm
{\it  Rudolf Peierls Centre for Theoretical Physics,\\
	1 Keble Road, Oxford, OX1 3NP, UK\\}

\vskip 1truecm

\end{center}

\centerline{\bf Abstract}
\medskip

Cachazo and Skinner have recently conjectured a new formula for the complete tree-level S-matrix of $\cN=8$ supergravity. The formula is an integral over the moduli space of rational curves in supertwistor space, and remarkably, manifests the full permutation symmetry of graviton scattering amplitudes. We check that the formula is parity invariance and reproduces the correct universal soft behaviour of graviton amplitudes.

\newpage


\section{Introduction}

Cachazo and Skinner~\cite{Cachazo:2012kg} have recently conjectured a new formula for the complete tree-level S-matrix of $\cN=8$ supergravity, extending a formula of Hodges~\cite{Hodges:2012ym,Hodges:2011wm} beyond the MHV sector. A remarkable feature of the new formulae is the ability to manifest the full permutation symmetry of gravitational scattering amplitudes, something that had not been achieved in previous expressions~\cite{Berends:1988zp,Mason:2008jy,Nguyen:2009jk}. Furthermore, the formula of Cachazo and Skinner is an integral over rational curves in supertwistor space, with the exciting prospect that there exists an underlying string theory description, such as the connected prescription for planar $\cN=4$ gauge theory amplitudes~\cite{Witten:2003nn,Roiban:2004yf}. 

\medskip

In this note, we consider two important checks of the Cachazo-Skinner formula. Firstly, we show that the formula is invariant under parity conjugation, which is highly non-obvious, since it interchanges rational curves of different degrees. Secondly, we demonstrate that the formula reproduces the universal soft behaviour of gravitational scattering amplitudes~\cite{Weinberg:1965nx}. This has been shown for MHV amplitudes since Hodges' formula obeys an `inverse soft' recursion relation, which manifests the correct soft behaviour~\cite{Hodges:2012ym}. However, soft limits provide another important check that the Cachazo-Skinner formula is working for all N$^k$MHV amplitudes.


\section{The Cachazo-Skinner Formula}

The formula
 \be
\cM_{n,d} = \int\frac{\prod_{r=0}^d d^{4|8}\cY_r}{\mathrm{vol}\,\mathrm{GL}(2,\mathbb{C})} \; \mbox{det}'(\widetilde\Phi)\;  \mbox{det}'(\Phi)\; \prod_{i=1}^{n} d^2\sigma_i \,\delta^2(\lb_i-\lb(\sigma_i))\,\exp \llbracket\, \mu(\sigma_i)\tlb_i\, \rrbracket 
\ee
is an integral over the moduli space of holomorphic maps $\cZ:\Sigma\to\CP^{3|8}$ from the Riemann sphere $\Sigma$ into $\cN=8$ supertwistor space $\CP^{3|8}$. Holomorphic maps of degree $d$ correspond to N$^k$MHV scattering amplitudes with $d=k+1$. The map is expanded in a basis of degree $d$ polynomials
\be
\cZ(\sigma) = \sum\limits_{j=0}^d \cY_r \, (\sigma^{\underline{1}})^r(\sigma^{\underline{2}})^{d-r}
\ee
in the homogeneous coordinates $\sigma^{\underline{\al}}=(\sigma^{\underline{1}},\sigma^{\underline{2}})$ on $\Sigma$ and the supertwistors $\cY_r$ are then coordinates on the moduli space of such holomorphic maps. The holomorphic form $\prod_r d^{4|8}\cY_r$ (modulo the $\mbox{GL}(1,\mathbb{C})$ rescaling of the supertwistors $\cY_r$) is not invariant under a change of basis of polynomials, however, this is compensated by the remaining parts of the integrand and gives a well defined integral on the moduli space.

\medskip

The vertex operators
\be
\delta^2( \lambda_i - \lambda(\sigma_i) ) \, \exp \llbracket\, \mu(\sigma_i)\tlb_i\, \rrbracket
\ee
are twistor space representatives of momentum eigenstate wavefunctions, described by the spinors $\{\lb_i,\tlb_i\}$. Here the supertwistor coordinates are $\cZ=(\lambda_\al,\mu^{\dal},\chi^a)$ and the notation $\llbracket \mu \tlb \rrbracket = \mu^{\dal}\tlb+\chi^a\teta_a$ includes the fermionic components. The insertions points are integrated over the Riemann sphere modulo SL$(2,\mathbb{C})$.

\medskip

The determinant $\det'(\widetilde\Phi)$ is formed from an $n\times n$ matrix with components
\be
\begin{aligned}
\widetilde\Phi_{ij}  & = \frac{[i,j]}{(i,j)}\; , \;\; i\neq j \\
\widetilde\Phi_{ii}  & = - \sum\limits_{j\neq i} \widetilde\Phi_{ij} \prod\limits_{r=0}^d \frac{(j,p_r)}{(i,p_r)}
\end{aligned}
\ee
where $\{p_0,p_1,\ldots,p_{d}\}$ are reference points on the Riemann sphere and $(i,j)=\epsilon_{\underline{\al}\underline{\beta}}\sigma_i^{\underline{\al}}\sigma_j^{\underline{\beta}}$ is the SL$(2,\mathbb{C})$-invariant form on the homogeneous coordinates. The components are independent of the choice of reference points on the support of delta-functions arising from the moduli space integral. The matrix $\widetilde{\Phi}$ has rank $(n\!-\!d\!-\!2)$ and hence a non-zero determinant is obtained by deleting rows $\cA=\{a_1,\ldots,a_{d+2}\}$ and columns $\cB=\{b_1,\ldots,b_{d+2}\}$ to form the minor $|\widetilde\Phi|^{\cA}_{\cB}$. Now consider the ratio
\be
\mbox{det}'(\widetilde\Phi) = \frac{| \widetilde\Phi |^{\cA}_{\cB}}{v_{\cA}\,v_{\cB}}
\ee
where 
\be
v_{\cA} = \sum\limits_{\substack{i,j\in\cA,\\ i<j} } (i,j)
\ee
is the Vandermonde determinant formed from elements of $\cA$. This is independent of the choice of $\cA$ and $\cB$ and hence fully permutation symmetric under interchange.

\medskip
 
Similarly, the determinant $\det'(\Phi)$ is formed from an $n\times n$ matrix with components
\be
\begin{aligned}
\Phi_{ij}  & = \frac{\la i,j\ra }{(i,j)}\; ,  \;\; i\neq j \\
\Phi_{ii}  & = - \sum\limits_{j\neq i} \Phi_{ij} \frac{\prod_{k\neq i} (i,k)}{\prod_{l\neq j}(j,l)} \prod\limits_{m=0}^{n-d-2} \frac{(j,p_m)}{(i,p_m)}
\label{psidef}
\end{aligned}
\ee
where now there are $(n\!-\!d\!-\!1)$ reference points. The components are again independent of this choice on the support of delta-functions in the measure. The matrix $\Phi$ has rank $d$ and hence we obtain a non-zero determinant $|\Phi|_{\cC,\cD}$ by deleting rows $\cC=\{c_1,\ldots,c_{n-d}\}$ and columns $\cD=\{d_1,\ldots,d_{n-d}\}$. Now the ratio
\be
{\det}'(\Phi) = \frac{|\Phi|^{\cC}_{\cD}}{v_{\cN \setminus \cC}\,v_{\cN\setminus\cD}}
\ee
is independent of $\cC$ and $\cD$ and hence fully permutation symmetric. Here the Vandermonde determinants are constructed from the complements of the sets $\cC$ and $\cD$ in $\cN=\{1,\ldots,n\}$, that is, from the rows and columns that have not been deleted.


\section{Parity Invariance}

Parity conjugation exchanges left-handed and right-handed spinors $|i\ra \leftrightarrow |i]$ and $+$ and $-$ helicity states, and thus sends an $n$-particle N$^k$MHV amplitude to an $n$-particle N$^{n-k-4}$MHV amplitude. For superamplitudes in $\cN=8$ supergravity, we have
\be
\cM_{n,d}\big(\lb_i,\tlb_i,\teta_i \big) = \int \prod_{i=1}^n d^{8}\eta_i \, e^{\left(i \sum\limits_{i=1}^n \teta_{i} \cdot \eta_i \right)}\, \cM_{n,n-d-2}\big(\tlb_i,\lb_i,\eta_i \big)\, .
\label{parity}
\ee
The correctness of this statement requires exchanging integrals over the moduli spaces of rational curves of degree $d$ and $(n\!-\!d\!-\!2)$, and therefore provides a highly non-trivial check of the formula.  

\medskip

In order to demonstrate invariance under parity conjugation we choose a local complex coordinate $z$ on the Riemann sphere, related to the homogeneous coordinates by a complex rescaling $\sigma^{\underline{\al}} = t^{1/d}(1, z)$ so that, for example
\be
 \qquad (i,j) = t_i^{1/d}t_j^{1/d}(z_i- z_j)\, .
\ee
In addition, we will perform the integrals over the $(\mu^{\dal},\chi^a)$-components of the supertwistor moduli $\cY_r$, against the exponentials in the vertex operators. In fact all fractional powers of $t_j$'s then cancel between components of the integrand, and the Cachazo-Skinner formula becomes
\be
\begin{aligned}
\cM_{n,d} = & \int \frac{\prod_{r=0}^d d^{2}\cY_r}{\mathrm{vol}\,\mathrm{GL}(2)}\; \prod\limits_{i=1}^n \frac{dt_i d z_i}{t_i} \;  \mbox{det}'(\widetilde\Psi) \;  \mbox{det}'(\Psi) \\
& \hspace{1cm}\times \prod_{i=1}^{n}  \,\delta^2\left( \lb_i-t_i\sum\limits_{r=0}^d \cY_r  z_i^r \right)\; \prod\limits_{r=0}^d \delta^{2|4}\left(\sum_{j=1}^n t_j z_j^r\tilde\lb_j \right)\, .
\label{CSnewvariables}
\end{aligned}
\ee
Here I have defined new matrices by
\be
\begin{aligned}
\widetilde\Psi_{ij} & = \frac{[i,j]}{( z_i- z_j)} \qquad i\neq j\\
\widetilde\Psi_{ii} & = -\sum\limits_{j\neq i} \frac{t_j}{t_i} \widetilde\Psi_{ij} \prod\limits_{r=0}^d\frac{( z_j-w_r)}{( z_i-w_r)}
\end{aligned}
\label{newmatrix1}
\ee
and similarly
\be
\begin{aligned}
\Psi_{ij} & = \frac{\la i,j\ra}{( z_i- z_j)} \\
\Psi_{ii} & = - \sum\limits_{j\neq i} \frac{t_i}{t_j} \Psi_{ij} \frac{\prod\limits_{k\neq i}( z_i- z_k)}{\prod\limits_{l\neq j}( z_j- z_l)} \prod\limits_{m=0}^{n-d-2} \frac{( z_j-w_m)}{( z_i-w_m)}\, .
\end{aligned}
\label{newmatrix2}
\ee
where the reference points are now denoted by $w_r$. The determinants $\mbox{det}'(\widetilde\Psi)$ and $\mbox{det}'(\Psi)$ are defined in exactly the same way as before, except that we now define
\be
|v_{\cA}| = \prod_{i\in\cA}t_i \prod\limits_{\substack{i,j\in\cA \\ i<j}}( z_i- z_j)\, .
\label{newvandermonde}
\ee
The determinants remain independent of the choices made on the support of delta-functions in equation~\eqref{CSnewvariables}.

\medskip

Since parity conjugation interchanges the brackets $\la\cdot,\cdot\ra \leftrightarrow [\cdot,\cdot]$, we would like to find a coordinate transformation that interchanges the components $\widetilde\Psi_{ij} \leftrightarrow \Psi_{ij}$ together with the degrees $d\leftrightarrow(n\!-\!d\!-\!2)$. Equations~\eqref{newmatrix1} and~\eqref{newmatrix2} immediately suggest the following change of coordinates
\be
\tilde t_i = \frac{1}{t_i \prod\limits_{j\neq i}( z_i- z_j)}\, .
\label{transf}
\ee
In the new coordinates we have
\be
\begin{aligned}
\widetilde\Psi_{ij} &= \frac{[ i,j ]}{( z_i- z_j )} \qquad i\neq j\\
 \qquad \widetilde\Psi_{ii} &= -\sum\limits_{j\neq i} \, \frac{\tilde{t}_j}{\tilde{t}_i}\, \widetilde\Psi_{ij}\, \frac{\prod\limits_{k\neq i}( z_i- z_k)}{\prod\limits_{l\neq j}( z_j- z_l)}  \prod\limits_{r=0}^d\frac{( z_j-p_r)}{( z_i-p_r)}
 \end{aligned}
\ee
and also
\be
\begin{aligned}
\Psi_{ij} & = \frac{\la i,j\ra}{( z_i- z_j)} \\
\qquad \Psi_{ii} & = - \sum\limits_{j\neq i} \, \frac{\tilde{t}_i}{\tilde{t}_j}\,  \Psi_{ij}\,   \prod\limits_{m=0}^{n-d-2} \frac{( z_j-p_m)}{( z_i-p_m)}\, .
\end{aligned}
\ee
Thus the minor $|\widetilde{\Psi}|_{\cA,\cB}$ is transformed into the minor $|\Psi|_{\cA,\cB}$ with the replacements $\la\cdot,\cdot\ra \rightarrow [\cdot,\cdot]$ and $d\rightarrow(n\!-\!d\!-\!2)$, and likewise the other way around.

\medskip


Now consider how the Vandermonde determinants $|v_{\cA}|$ transform. After some straightforward combinatorics we find that
\be
|v_{\cA}| = |\tilde{v}_{\cN\setminus\cA}| \times 
\Big( \prod\limits_{i<j}( z_i- z_j) \prod\limits_{i=1}^n\tilde{t}_i\; \Big)^{-1}
\ee
where the new Vandermonde determinant $|\tilde{v}_{\cN\setminus\cA}|$ is defined as in equation~\eqref{newvandermonde} with respect to the transformed variables $\tilde{t}_j$ and the complement of the set $\cA$ in $\cN=\{1,\ldots,n\}$. The products of Vandermonde determinants $\tilde{v}_{\cN\setminus\cA}\,\tilde{v}_{\cN\setminus\cB}$ and $\tilde{v}_{\cC}\,\tilde{v}_{\cD}$ now combine with the minors to form the new determinants $\det'(\Psi)$ and $\det'(\widetilde{\Psi})$ of the parity conjugated formula, modulo an overall factor of
\be
\Big( \prod\limits_{i<j}( z_i\!- \!z_j) \prod\limits_{i=1}^n\tilde{t}_i\; \Big)^4\, .
\label{vd1}
\ee
However, we will now see that this is compensated by the transformation of the bosonic and fermionic delta-functions. 

\medskip

The coordinate transformation~\eqref{transf} is the same as that used in proving parity conjugation invariance of the connected prescription of twistor-string theory for tree-level amplitudes in planar $\cN=4$ gauge theory~\cite{Roiban:2004yf,Witten:2004cp}. There it was shown that the same coordinate transformation~\eqref{transf} indeed transforms the remaining delta-functions and moduli integrals from curves of degree $d$ to curves of degree $(n\!-\!d\!-\!2)$. Concretely, they showed that
\begin{multline}
\int \prod\limits_{r=1}^d d^2\cY_r \prod\limits_{i=1}^n \delta^2\left( \lb_i-t_i\sum_{r=0}^d \cY_r  z_i^r \right) \prod\limits_{r=0}^d \delta^2\left( \sum\limits_{i=1}^n t_i\,  z_i^r \, \tlb_i  \right) \\
=\prod\limits_{m=0}^{n-d-2} d^{2}\tilde{\cY}_m \prod\limits_{i=1}^n \delta^2\left( \tlb_i - \tilde{t}_i \sum\limits_{m=0}^{n-d-2} \tilde{\cY}_m\,  z_i^{\,m}  \right) \prod_{m=0}^{n-d-2}\delta^2\left( \sum\limits_{i=1}^n\tilde{t}_i\,  z_i^{\,m}\, \tlb_i \right)  \\
\times   \Big( \;\prod_{i<j}( z_i -  z_j) \prod\limits_{j=1}^n \tilde{t}_j \Big)^{4}\, ,
\label{vd2}
\end{multline}
where the new variables $\tilde{\cY}_m$ are coordinates on the moduli space of degree $(n\!-\!d\!-\!2)$ maps. The second line has exactly the same form as the first with the replacements $\lb_i\leftrightarrow\tlb_i$ and $d\leftrightarrow(n-d-2)$. 

\medskip

Finally, we can express the fermionic delta-functions using an inverse Fourier transform
\be
\begin{aligned}
\prod\limits_{r=0}^d \delta^{0|8}\left( \sum\limits_{i=1}^n t_i\,  z_i^r \, \teta_i  \right) = \int \prod_{i=1}^n d^{8}\eta_i \, e^{\left(i \sum\limits_{i=1}^n \teta_{i} \cdot \eta_i \right)} \, \prod\limits_{m=0}^{n-d-2} \delta^{0|8}\left( \sum_{i=1}^n \tilde{t}_i\, \tilde z_i^{\, m} \eta_i \right) \\ 
\times \Big( \;\prod_{i<j}( z_i -  z_j) \prod\limits_{j=1}^n \tilde{t}_j \Big)^{-8} \, .
\label{vd3}
\end{aligned}
\ee
The Vandermonde determinant factors in equations~\eqref{vd1},~\eqref{vd2} and~\eqref{vd3} now cancel between various components of the formula. Thus, taking all components into account, we have shown that the Cachazo-Skinner formula is indeed invariant under parity conjugation, as expressed in equation~\eqref{parity}.


\section{Soft Limits}

\begin{figure}[htp]
\centering
\includegraphics[height=2.5cm]{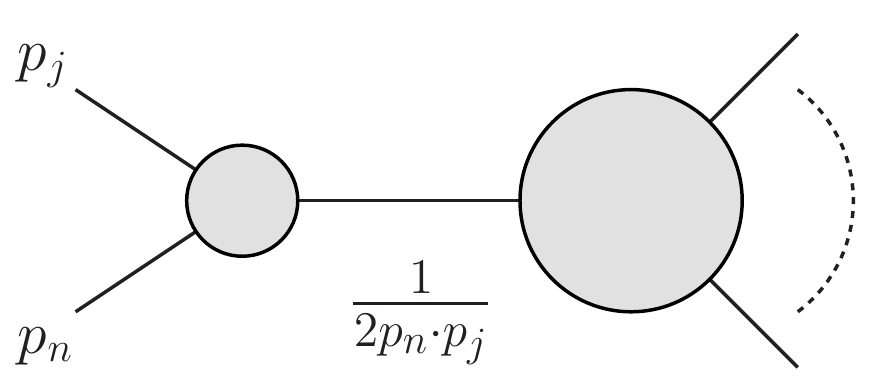}
\caption{\emph{Feynman diagrams contributing to the soft limit.}}
\label{figure:soft}
\end{figure}

Gravitational scattering amplitudes have singularities when the momentum of a graviton becomes soft $p_n\to0$, which arise from propagators of the form $1/ p_n\cdot p_j$ in the Feynman diagram expansion - see figure~\ref{figure:soft}. The behaviour of gravitational amplitudes in this limit are determined by the factorisation properties and have a universal form; for soft gravitons of positive helicity
\be
\begin{aligned}
\cM_{n,d}  \longrightarrow & \sum\limits_{j=1}^{n-1} \left( \frac{\la \zeta, j\ra}{\la \zeta, n\ra}[n,j] \right)^2 \frac{1}{2p_n\cdot p_j} \; \cM_{n-1,d} \\
& \; = \sum\limits_{j=1}^{n-1} \frac{[n,j]}{\la n, j\ra} \frac{\la \zeta, j\ra^2}{\la \zeta, n \ra^2} \; \cM_{n-1,d}
\label{softfactor}
\end{aligned}
\ee
where $\zeta^\al$ is an auxiliary spinor~\cite{Weinberg:1965nx,Berends:1988zp}. For gravitons of negative helicity, the soft behaviour is obtained by parity conjugation of equation~\eqref{softfactor}. Since we have proven invariance under parity conjugation above, we concentrate here on the simpler positive helicity case, which leaves the degree $d$ of the curve unchanged.

In order to prove that the Cachazo-Skinner formula reproduces the soft behaviour~\eqref{softfactor}, we must integrate over the soft graviton $n$. For a positive helicity graviton we set $\tilde\eta_n=0$. We then identify $1/ \la n, j\ra$ singularities in the integral over the position of the soft graviton vertex operator, and determine whether the soft divergent terms reproduce the correct behaviour~\eqref{softfactor}.

\medskip

We first consider the determinant $\det'(\Psi)$. Since the matrix $\Psi$ has rank $d$, it is sensible to remove rows and columns that include the soft graviton $n$. The only place that $n$ then appears in $\det'(\Psi)$ is through the diagonal elements $\tPsi_{ii}$ where $i\neq n$,
\be
\begin{aligned}
\Psi_{ii} &=  \frac{t_i}{t_n}\frac{\la i, n \ra}{(z_i -  z_n)} \frac{( z_i- z_n)\prod\limits_{k\neq i,n}( z_i -  z_k)}{\prod\limits_{l\neq n}( z_j -  z_l)} \prod_{m=0}^{n-d-2} \frac{( z_n -w_m)}{( z_i - w_m)} \\
&+\sum_{j\neq i,n} \frac{t_i}{t_j} \frac{\la i, j \ra}{( z_i -  z_j)}\, \frac{( z_i- z_n)\prod\limits_{k\neq i,n}( z_i- z_k)}{( z_j- z_n)\prod\limits_{l\neq j,s}( z_j- z_l)} \, \prod_{m=0}^{n-d-2} \frac{( z_j - w_m)}{( z_i - w_m)}\, .
\label{PsiExpand}
\end{aligned}
\ee
Now after integrating out the soft graviton $n$, we expect a formula involving only $(n-d-3)$ reference points $w_r$.  Hence let us choose one reference point to be $ z_n$. Then the first term in equation~\eqref{PsiExpand} vanishes, and the dependence on $z_n$ cancels out in the remaining terms, leaving
\be
\begin{aligned}
\Psi_{ii} = \sum_{j\neq i,n} \frac{t_i}{t_j} \frac{\la i j \ra}{ z_i- z_j}\, \frac{\prod\limits_{k\neq i}( z_i- z_k)}{\prod\limits_{l\neq j}( z_j- z_l)} \, \prod_{m=0}^{n-d-3} \frac{( z_j - w_m)}{( z_i- w_m)} \, .
\end{aligned}
\ee
This equation defines the diagonal elements of the corresponding matrix $\Psi_{ij}$ associated to the remaining particles $\{1,\ldots,n-1\}$. Hence the determinant $\det'(\Psi)$ reduces immediately to the corresponding determinant for the smaller set $\{1,\ldots,n-1\}$ and we have removed any dependence on the soft graviton. 

\medskip

The determinant $\det'(\tPsi)$ certainly does depend on the soft graviton, and is the origin of the universal soft factor. Since the matrix $\tPsi$ has rank $(n\!-\!d\!-\!2)$, and it is now important to remove rows and columns that do not include $n$. Here we will choose to remove from $\tPsi$ rows and columns corresponding to the particles $\{n-d-2,\ldots,n-1\}$. The remaining maximal rank matrix will be denoted by $\tP$ and we will use the notation $|\,\tP\,|^{a_1\ldots a_i}_{b_1\ldots b_j}$ for the minor that is obtained by deleting rows $\{ a_1,\ldots,a_i\}$ and columns $\{b_1,\ldots,b_j\}$ from the matrix $\tP$.

\medskip

The dependence on $n$ in $\det'(\tPsi)$ comes from three sources
\be
\begin{aligned}
\tP_{ni} & = \frac{[n,i]}{z_n-z_i} \\
\widetilde{\cP}_{nn} & = - \sum_{i=1}^{n-1} \frac{t_i}{t_n}\, \tP_{nj} \prod\limits_{r=0}^d \,\frac{z_i-w_r}{z_n-w_r} \\
\tP_{ii} & = \tQ_{ii} - \frac{t_n}{t_i}\, \tP_{ni} \prod\limits_{r=0}^d \frac{z_n-w_r}{z_i-w_r}
\end{aligned}
\ee
where in the last equation we have isolated the $n$-dependence in the second term. Here we have introduced the notation $\tQ$ for the maximal rank matrix obtained by removing the rows and columns $\{n-d-2,\ldots,n-1\}$ from the $\tPsi$-matrix for the remaining particles $\{1,\ldots,n-1\}$. 

\medskip

We can now systematically expand the determinant $|\,\tP\,|$ in powers of the square brackets $[n,\cdot\,]$ involving the soft graviton. For example, we can first expand out the determinant as follows
\be
|\,\tP\,| = \tP_{nn}\, | \, \tP\, |^n_n + \sum_{i=1}^{n-d-3} \sum_{j=1}^{n-d-3}(-1)^{i+j+1}\, \tP_{ni}\, \tP_{nj} \, |\,\tP\,|^{ni}_{nj}\, .
\label{detexpand}
\ee
The components of matrix $\tQ$ differ from those $\tP$ only along the diagonals. Hence, in order to make the dependence on the soft graviton completely explicit, we can further expand the minor
\be
 |\,\tP\,|^n_n = |\,\tQ\,| + 
 \sum_{i=1}^{n-d-3} \sum\limits_{1\leq a_1 < \ldots < a_i\leq n-d-3} \; \prod\limits_{j=1}^i \, \left( - \frac{t_n}{t_{a_j}}\tP_{na_j} \prod\limits_{r=0}^d\frac{z_n\!-\!w_r}{z_{a_j}\!-\!w_r}\right) \, |\, \tQ \, | ^{a_1\ldots a_i}_{a_1\ldots a_i}
\ee
and similarly for the minor $|\,\tP\,|^{ni}_{nj}$. 

\medskip

We will not require the full details of this expansion in order to understand the soft behaviour of Cachazo-Skinner formula. The important observation is that the expansion in square brackets $[n,\cdot\,]$ leads to three kinds of term only:
\begin{enumerate}
\item $[n,j]$. \\ These terms reproduce the correct universal soft behaviour.
\item $[n,j_1][n,j_2]\cdots$, $j_1\neq j_2 \cdots$. \\ These terms are subleading in the soft limit.
\item $[n,j_1]^2[n,j_2][n,j_3]\cdots$ with $j_2\neq j_3\cdots$. \\ These terms cancel in pairs before the soft limit is taken.
\end{enumerate}
We now consider each of them in turn.

\medskip

1. Terms with a single square bracket arise from the term $\tP_{nn} | \, \tQ\, |$ in the expansion of the determinant $| \, \tP \, |$, that is
\be
- \sum_{i=1}^{n-1} \left( \frac{t_i}{t_n}\, \frac{[n,i]}{z_n-z_i} \prod\limits_{r=0}^d \,\frac{z_i-w_r}{z_n-w_r} \right) |\, \tQ \, |\, .
\ee
Note that $| \, \tQ \, |$ is the minor required to form the determinant $\det'(\tPsi)$ for the remaining particles $\{1,\ldots,n-1\}$. Hence we would expect the universal soft factor to arise from these terms. Therefore our task is to isolate terms with poles as $\la n,j\ra\to0$ in the integral
\be
\int \frac{dt_n}{t_n} d z_n  \left( \frac{t_j}{t_n}\frac{[n,j]}{z_n-z_j} \prod_{r=0}^d \frac{z_j-w_r}{z_n-w_r}     \right) \delta^2(\lambda_n - t_n \lb( z_n))\, .
\ee
and examine their soft behaviour. The argument of the delta function has many complex roots, which must all be included to evaluate the integral in full. However, we will adapt an argument of~\cite{Roiban:2004yf} which shows that only one root gives rise to a pole when $\la n,j\ra\to0$, and that this may be straightforwardly evaluated.

\medskip

We first perform $t_n$-integral using one of the $\delta$-functions,
\be
\int d z_n  \left(  t_j \frac{[n,j]}{z_n-z_j} \prod_{r=0}^d \frac{z_j-w_r}{z_n-w_r}  \right) \frac{\lb^1( z_n)}{(\lb_n^1)^3} \delta\left( \frac{\lb_n^2}{\lb_n^1} - \frac{\lb^2( z_n)}{\lb^1( z_n)}  \right)\, .
\ee
On the support of the $\delta$-functions for particle $j$ we are free to subtract from the argument of the $\delta$-function the corresponding expression $\lb^2_j/\lb^1_j-\lb^2(z_j)/\lb^1(z_j)$, so that the argument of the $\delta$-function becomes
\be
\delta  \left( \frac{\la nj \ra}{\lb_n^1\lb_j^1} - \left[ \frac{\lb^2( z_j)}{\lb^1( z_j)} - \frac{\lb^2( z_n)}{\lb^1( z_n)} \right] \right) \, .
\ee
The term in square brackets is a rational function of $z_n$, which vanishes when $ z_n \to z_j$. Hence, introducing the new variable $\omega=z_n-z_j$, it may be expressed $\omega\, F(\omega,z_j)$ for some other rational function $F$. Thus we are considering an integral of the form
\be
\int \frac{d\omega}{\omega} f(\omega) \, \delta\left( \frac{\la n,j \ra}{\lb_n^1\lb_j^1} - \omega F(\omega,z_j)\right)\, .
\ee
where the rational function $f(z)$ is regular away from the reference points. We now consider the roots of the argument of the $\delta$-function in the limit $\la n,j \ra\to0$, following the argument of~\cite{Roiban:2004yf}. There is one root where $\omega$ is becoming small (the same order as the angle  bracket $\la n,j \ra$) and where the function $F(\omega,z)$ is of order unity. In addition, there may be other roots where $F(\omega,z)$ is becoming small. However, performing the integral against the $\delta$-function produces a factor $1/g'(\omega)$, which is of order unity when evaluated on any of the roots. Thus, provided the reference points are generic, the only pole of the form $1/\la n,j\ra$ comes from the factor of $1/\omega$ evaluated on the small root.

\medskip

In computing the contribution from this root in the limit $\la n,j\ra\to0$, the dependence on the function $F(\omega, z)$ cancels, we can set $ z_s= z_j$ in the remaining integrand. The important point here is that the dependence on the reference points $w_r$ cancels when $z_n=z_j$. We find 
\be
 \sum_{j=1}^{n-1}\,  [n,j] \, \frac{t_j\lb^1(\sigma_j)}{(\lb_s^1)^3}\,  \frac{\lb_s^1\lb_j^1}{\la s,j\ra} = \sum_{j\neq s} \frac{[s,j]}{\la s,j\ra} \frac{\la j,\zeta\ra^2}{\la s ,\zeta \ra^2}\, ,
\ee
where we have introduced the spinor $\zeta^{\al}$ which extracts the first component $\lb^1_j = \la j,\zeta\ra$. However, momentum conservation allows this to be changed to any other auxiliary spinor. This is the universal gravitational soft factor.

\medskip

2. Terms with higher numbers of square brackets $[n,j_1]\ldots[n,j_m]$ arise from both terms in the expansion~\eqref{detexpand} of the determinant $|\,\tP\,|$. Here we consider the case where none of the square brackets are repeated: $j_1\neq j_2 \neq \cdots$. These terms do lead to simple poles $1/\la n,j_1\ra$ when the soft graviton is integrated out. However, we will now show that they are subleading in the soft limit.

\medskip

The dependence on the soft graviton $n$ of each term of this kind is contained an integral of the following form
\be
\int \frac{dt_n}{t^m_n} \int dz_n\, f(z_n) \left( \frac{[n,j_1]}{z_n-z_{j_1}} \ldots \frac{[n,j_m]}{z_n-z_{j_m}}  \right) \delta^2(\lb_n-z_n\lb(z_n))
\ee
where $f(z)$ is a polynomial function with zeroes only at the reference points $w_r$. Consider for simplicity the terms with two square brackets, where the function $f(z)=1$. Performing the $t_n$ integral and expanding in partial fractions, we have
\be
\frac{1}{z_i-z_j}\int dz_n\,  \left( \frac{1}{z_n-z_i}-\frac{1}{z_n-z_j} \right) \frac{[n,i][n,j]}{(\lb_n^1)^2\,\lb^1(z_n)} \delta\left( \frac{\lb_n^2}{\lb_n^1} - \frac{\lb^2( z_n)}{\lb^1( z_n)} \right)
\ee
We now apply the same argument as above to extract the contribution from this integral leading to poles as $\la n, j\ra\to0$, with the result  
\be
\begin{aligned}
\frac{[n,i]\,[n,j]}{\la n,i\ra\la n,j \ra} \frac{\la i,j\ra}{z_i-z_j}\ .
\end{aligned}
\ee
which is subleading in the soft limit $p_n\to0$. 

\medskip

For terms with $m$ square brackets, the presence of the polynomial function $f(z)$ does not change the argument. Suppose that the soft graviton momentum is of order $\cO(\epsilon)$ with $\epsilon\to0$. Then the square brackets in the numerator always contribute $\cO(\epsilon^{m/2})$. As a consequence of the factor $1/t_n^m$ in the measure, the integration over $t_n$ contributes another $\cO(\epsilon^{m/2-2})$. The total scaling $\cO(\epsilon^{m-2})$ means that all such terms are subleading in the soft limit.

\medskip

3. Terms with squared square brackets $[n,j]^2$ contain potential additional soft divergent terms. However, all such terms cancel in pairs in the expansion of the determinant $|\,\tP\,|$ before any soft limit is approached. Firstly, there are terms from the expansion of $\tP_{nn}|\tP|^n_n$, which are
\be
\begin{aligned}
\sum_{j=1}^{n-d-3} \left(- \frac{t_j}{t_n}\, \tP_{nj} \prod\limits_{r=0}^d \,\frac{z_j-w_r}{z_n-w_r} \right) & \left( - \frac{t_n}{t_j}\tP_{nj} \prod\limits_{r=0}^d\frac{z_n\!-\!w_r}{z_j\!-\!w_r}\right) |\, \tP\, |^{nj}_{nj} \\
& \hspace{2.5cm}  =  \sum_{j=1}^{n-d-3} \left( \frac{[n,j]}{z_n-z_j} \right)^2|\, \tP\, |^{nj}_{nj}\, .
\end{aligned}
\ee
Secondly, there are terms from $\sum_i\sum_j\tP_{ni} \tP_{nj} |\, \tP\, |^{ni}_{nj}$, which take the same form but appear with the opposite sign
\be
- \sum\limits_{j=1}^{n-d-3} \left(\frac{[n,j]}{z_n-z_j}\right)^2 |\,\tP\,|_{nj}^{nj}\, .
\ee
Hence all squared square brackets cancel out in pairs and cannot contribute additional soft divergences.

\medskip

This completes the argument that the Cachazo-Skinner formula reproduces the correct universal soft behaviour expected for all N$^k$MHV amplitudes.

\section{Conclusions}

The Cachazo-Skinner formula is a remarkable conjecture for all tree-level scattering amplitudes of $\cN=8$ supergravity. In this paper, we have shown that the formula is parity invariant and reproduces the universal soft behaviour of gravitational amplitudes. These provide non-trivial evidence that the formula is working beyond the MHV sector. 

\medskip

Further evidence could be provided by examining collinear limits, which are more subtle for gravitational amplitudes~\cite{Bern:1998sv}, and more general multi-particle factorisation properties, perhaps using the methods of~\cite{Skinner:2010cz}. This could indeed lead to a proof of the Cachazo-Skinner formula by on-shell recursion methods, as has been demonstrated for Hodges' formula for MHV amplitudes~\cite{Hodges:2012ym}.

\section*{Acknowledgements}

I would like to thank Tim Adamo for useful discussions and comments on the draft. I am supported by an STFC Postgraduate Studentship.


\bibliographystyle{JHEP}
\bibliography{Gravity}

\end{document}